\documentclass[12pt,a4paper]{article}
\usepackage{amssymb}

 \newcommand{\bm}[1]{\mbox{\boldmath $#1$}}

 \newcommand{\NN}{\mathbb N}
 \newcommand{\RR}{\mathbb R}
 
 \newcommand{\ZZ}{\mathbb Z}

\newtheorem{defin}{Definition}
\newtheorem{theorem}{Theorem}

\newtheorem{prop}{Proposition}
\newtheorem{coro}{Corollary}

 \newcommand{\be}{\begin{equation}}
 \newcommand{\ee}{\end{equation}}

 \input{epsf}
 
\begin{document}
 \bibliographystyle{unsrt}

 \parindent0pt

\begin{center}
{\large \bf Diffractive Point Sets with Entropy}

 \end{center}
 \vspace{3mm}
 
 \begin{center}  {\sc Michael Baake}$^1$ and {\sc Robert V. Moody}$^2$ 
 
\vspace{5mm}

 1) Institut f\"ur Theoretische Physik, Universit\"at T\"ubingen, \\

 Auf der Morgenstelle 14, D-72076 T\"ubingen, Germany  \\

 2) Department of Mathematical Sciences, University of Alberta, \\

 Edmonton, Alberta T6G 2G1, Canada
 
\end{center}
 
\vspace{8mm}
\centerline{Dedicated to Hans-Ude Nissen on the occasion
                  of his 65th birthday}
\vspace{10mm}
 
\begin{abstract} After a brief historical survey, the paper introduces 
the notion of entropic model sets (cut and project sets), and, 
more generally, the notion of diffractive point sets with entropy.
Such sets may be thought of as generalizations of lattice gases.
We show that taking the site occupation of a model set stochastically 
results, with probabilistic certainty, in well-defined diffractive 
properties augmented by a constant diffuse background.
We discuss both the case of independent, but identically distributed
(i.i.d.) random variables and that of independent, but different 
(i.e., site dependent) random variables. Several examples are shown.

\end{abstract}

\parindent15pt
\vspace{15mm}

\subsection*{Introduction}

Diffraction is one of the most important ways of identifying long-range
order in mathematical and physical structures.
In this paper, we look at the effects on diffraction that occur
in certain periodic and quasiperiodic point sets when the occupation
of the point sites is taken stochastically rather than deterministically,
with independence between the different sites.
Under fairly mild assumptions, which are certainly
valid for lattices and model sets, we show that the effect is simply one
of scaling down the diffraction pattern by a constant factor and adding in a
constant background. In the case of lattices, this type of phenomenon
is well known \cite{Cowley, Welberry}. 
What is new here is that it remains true for a large
class of non-periodic structures (Theorems 1 and 2) and, in particular,
for all regular model sets (also called cut and project sets).

The results are, on the one hand, a suitable reminder of the difficulty
of interpreting the meaning of diffractivity, and, on the other hand,
of the robustness of diffractivity under certain deformations
and modifications of the underlying set.

It might be interesting to quickly review the history of aperiodic order and
take a look at the reasons why stochastic forms of aperiodic structures
seem to be a natural extension beyond the world of strict perfection.
In the early eigthies, a new type of ordered state was found,
both experimentally \cite{Shechtman,INF} and theoretically
\cite{Kramer82,KN}. These discoveries, made independently of one another,
created an enormous amount of scientific activity because
the new ordered states, quickly dubbed quasicrystals, had properties 
previously thought to be incompatible with one another; namely,
long-range orientational order, strong enough to produce sharp diffraction
images, and at the same time non-crystallographic symmetries such as 
icosahedral \cite{Shechtman,KN} or twelvefold symmetry \cite{INF}.

Since nothing has ever been discovered for the first time, one might
expect precursors of this, and this is indeed correct. Clearly, Penrose's 
famous tiling of the plane
with fivefold symmetry was important, particularly when coupled with de
Bruijn's algebraic analysis that showed that it was also 
diffractive\footnote{If not explicitly specified, the proper references
are obvious by the names given, and can be found in \cite{SO}.}. Ammann
investigated this further and also found the matching icosahedral
tiling made from two rhombohedra. However, this was not generally known
in the physics community (a brief remark can be found in
\cite{Mackay}), and has never been published by him. Even his planar
results were published only much later \cite{Ammann}, though some results
are contained in \cite{GS}.

In fact, the history goes back quite a bit before this. In
the late thirties of this century, Kowalewski \cite{Kowa} investigated the
possibilities of filling Kepler's triacontahedron with the two rhombohedra
mentioned above, but apparently did not realize the possibility
of filling the entire space with them. 

Kepler himself was very much 
interested in space fillings in his time, and his famous plate of
planar tilings (resp.\ seeds of them) shows a considerable patch
of a tiling with pentagons, decagons and rhombi, see the first plate
in \cite{GS}. In modern terminology \cite{BS}, 
this would be in the same MLD-class as the famous
Penrose tiling, in the sense that there is a derivation rule with a radius
much smaller than the patch shown such that, on the size displayed, there is
no way to tell the two tilings apart. 
This is probably not an accident because Kepler was 
well aware of the problems of space fillings, as were other people before
him, such as D\"urer who constructed a mechanism to create a fivefold
twin made from pentagons and rhombi. 
Generally, the investigation
of geometric form was well on the way. D\"urer's polyhedron in
his ``melancholia'' has puzzled generations of scientists and art historians
-- with a really promising solution being found only very recently
by Hans-Ude Nissen \cite{Nissen}.

Coming back to this century, the development of the theory of
incommensurate structures by the Nijmegen group and the new
developments in the theory of quasicrystals showed that there is
a lot more to the geometry and symmetry of the solid state than
anticipated by ordinary school knowledge. Probably the most
puzzling aspect in the beginning was the combination of perfect
diffractivity (in the sense of a Bragg spectrum) with non-crystallographic
symmetry. But again, the final explanation, in terms of the projection
method \cite{KN}, had a precursor, this time in pure mathematics.

Harald Bohr, the younger brother of Niels, developed the theory
of quasi- and almost periodic functions in the twenties. The basic
idea was to describe non- but quasi-periodic functions as sections
through periodic functions in higher dimensions. In this sense, the
cut-and-project method owes a lot to his ideas. In the late
1960's, Yves Meyer studied the harmonic analysis of point sets in the
context of algebraic numbers. In the process he rediscovered cut
and project sets (here called {\em model sets}), though now in the much wider
setting of locally compact Abelian groups, and introduced a very important
class of ordered point sets, now called Meyer sets \cite{Meyer}
(see later for one characterization of Meyer sets).

After this historically motivated introduction, let us come to the
aim of this article. Although all the above mentioned connections
might indicate that the (quasi-) crystalline world is perfect, in the 
sense that the alloys displaying such diffraction images are, reality
tells us nowadays that this is not so \cite{NB,JRB}. 
In fact, quite early it was pointed
out by Elser that, in order to explain the stability of such alloys, 
one might need an {\em entropic} side of the picture, an idea that led the
Cornell group to develop the idea of a random tiling. From a more
mathematical point of view, this is not fully satisfying because quite
a number of questions concerning the diffractivity, and even the
well-definedness of some of the ensembles, are still unanswered.

This is the point we want to consider and start to develop. However,
we will not adopt the random tiling picture here, because it seems
not yet fully in reach for a rigorous treatment (compare \cite{Richard}
and references therein for some recent developments).
Instead, as an intermediate step, we shall rather consider a setup 
of ideal {\em model sets}, or even more general diffractive point sets,
that are coupled to  stochastic processes to thin them out. This way, we
can introduce some randomness into the picture, even with positive
entropy density, without losing control of the diffractivity. 
This can be seen as a generalization of the diffraction theory of
lattice gases which belongs to the standard body of literature,
see \cite{Guinier,Cowley,Welberry} and references therein.

In doing so, we will actually arrive at results that exactly meet
the expectation, but with the extra benefit of providing proofs
for them, i.e.\ making a good deal of folklore rigorous in this way.
It will turn out that there is a {\em natural} extension of the
diffraction theory of lattices gases beyond lattices, {\em provided}
one uses an approach that avoids techniques based upon translation
invariance. Although the actual methods and results employed from
probability theory and ergodic theory are pretty standard in 
mathematics, they are much less familiar to physicists.

Let us now briefly sketch how the article is organized.
We start with a section on the diffraction of lattice gases
(without interaction) and its connection to entropy density. 
This gives us the opportunity to review
some well-known results in a different setting that matches the
generalizations derived later. We hope that the reader can adjust
to our approach that way without too much pain. This is followed
by our general setup, where we introduce a rather general type
of point sets which are accessible with our methods. Common examples
such as model sets (or cut and project sets) are contained as
special cases.

The remainder of the article is then devoted to the diffraction
of point sets with independent stochastic occupation of sites.
First, the focus is on the situation of independent, but identically
distributed (i.i.d.) random variables, the case most frequently
studied. Theorem 1 gives the result for this case. This is illustrated
by some examples, and model sets in particular.

More general, and less obvious, is the treatment of independent,
but not necessarily identically distributed, random variables, which
leads to Theorem 2. Among the applications are weighted model sets
and their stochastic counterpart, and, more specifically, weighted
model sets where the weights are determined by a so-called invariant
density \cite{BM97.1,BM97}. This way, we are able to keep certain 
aspects of point and inflation symmetries. We believe that this
application is of particular value in the discussion of perfect
versus random tiling order, as it really is a first step of an
intermediate picture.

Our concluding remarks try to relate the results to other 
investigations and to point towards the next steps that should
be taken.

\subsection*{Diffraction from a lattice gas}

In order to keep things simple, and to familiarize the reader
with our approach, we start with the description of the lattice 
situation and give proper definitions for the general setup later.
Let $\Gamma$ be a lattice in $\RR^n$, i.e.\ a discrete Abelian subgroup
of $\RR^n$ such that $\RR^n/\Gamma$ is compact \cite{Schwarz}. Equivalently,
there are $n$ linearly independent vectors $\bm{b}_1, \ldots , \bm{b}_n$,
called the basis vectors of $\Gamma$,
so that $\Gamma = \ZZ\bm{b}_1 \oplus \cdots \oplus \ZZ\bm{b}_n$.
Since we will be talking about Fourier transforms, we will also
need the dual (or reciprocal) lattice of $\Gamma$, defined by
\be \label{dualdef}
    \Gamma^{\circ} \; := \; \{ x \in \RR^n \mid
        x \cdot y \in \ZZ \quad \mbox{for all} \quad y\in\Gamma \}
\ee
where $x \cdot y$ denotes the Euclidean scalar product.

Next, define {\em Dirac's comb} as the characteristic distribution
\be \label{mudef}
    \omega \; = \; \omega^{}_{\Gamma} \; := \; \sum_{x\in\Gamma} \delta_x
\ee
on $\Gamma$, where $\delta_x$ is Dirac's distribution at point $x$, i.e.\ 
\be
         (\delta_x \, , \phi) \; := \; \phi(x)
\ee
for all test functions $\phi$. In particular, one gets
\be
        (\omega , \phi) \; = \; \sum_{x\in\Gamma} \phi(x) 
\ee
which is well defined for all rapidly decreasing functions (Schwartz functions), 
hence $\omega^{}_{\Gamma}$ is a tempered distribution \cite{RS}.

To deal with diffraction, we need the corresponding autocorrelation
distribution, $\gamma^{}_{\omega}$, of $\Gamma$, also called its Patterson
function (although it is a distribution\footnote{It would actually be
slightly more appropriate to adopt the setup of measure theory, where
$\gamma^{}_{\omega}$ would represent a tempered measure, see \cite{Hof}
for this complementary approach.}). With the abbreviation
\be
      \Gamma_r \; := \; \Gamma \cap B_r(0)
                      \; = \;  \{ x \in \Gamma \mid |x| \leq r \} \, ,
\ee
$\gamma^{}_{\omega}$ can be defined and calculated as follows
\be
   \gamma^{}_{\omega} \; := \; \lim_{r\rightarrow\infty}
               \frac{1}{{\rm vol}(B_r(0))} \sum_{x,y \in \Gamma_r} \delta_{x-y}
               \; = \; d \cdot \omega
\ee
where $d$ is the density of $\Gamma$, i.e.\ the number of lattice points
per unit volume. 

By the Fourier transform of a Schwartz function $\phi$ we mean
\be \label{fourierdef}
      \hat{\phi} (k) \; := \; \int_{\RR^n} e^{-2\pi i k\cdot x} \, \phi(x) dx
\ee
which is again a Schwartz function \cite{RS}. The inverse operation is given by
\be \label{invfourierdef}
      \check{\psi} (x) \; = \; \int_{\RR^n} e^{2\pi i x\cdot k} \, \psi(k) dk \, .
\ee
This definition results in the usual properties, such as 
$\check{\hat{\phi}\,}=\phi$
and $\hat{\check{\psi}\,}=\psi$. The convolution theorem takes the form
$\widehat{\phi_1 * \phi_2} = \hat{\phi}_1 \cdot \hat{\phi}_2$ where convolution
is defined by
\begin{equation}
          \phi_1 * \phi_2 \, (x) \; = \; \int_{\RR^n} \phi_1(x-y)\phi_2(y)dy \, .
\end{equation}
Finally, the matching definition of the Fourier transform of a 
distribution $T$ is
\be
 ( \hat{T}, \phi ) \; := \; ( T, \hat{\phi} )
\ee
for all Schwartz functions $\phi$, as usual.

Now, the diffraction by the lattice $\Gamma$ is described by the
Fourier transform of its autocorrelation \cite{Cowley}, and in this
case we have $\hat{\gamma}^{}_{\Gamma} = d \cdot \hat{\omega}^{}_{\Gamma}$. 
To calculate the latter, we employ Poisson's summation formula for tempered
distributions, cf.\ p.\ 254 of \cite{Schwartz}, which reads
\be
         \widehat{\sum_{x\in\Gamma}\delta_x}
           \; = \; d \cdot \sum_{k\in\Gamma^{\circ}} \delta_k
\ee
and can easily be proved from the corresponding Poisson
summation formula for Schwartz functions. So we get
\be
         \hat{\gamma}^{}_{\omega} \; = \; d_{\;}^2 \cdot
                           \sum_{k\in\Gamma^{\circ}} \delta_k
\ee
which is the well-known result that the diffraction from
point scatterers of equal strength on the points of a lattice
is a pure point measure, consisting of  periodically placed
point measures on the dual lattice. Note that the strength
of the peak at $k=0$ is $d^2$, the square of the density of $\Gamma$,
as it must be.

Let us now move on to the corresponding lattice gas, i.e.\ to the
point set obtained from $\Gamma$ by removing points from it
stochastically. To describe this, we define a new measure $\omega^{}_s$ 
(with suffix $s$ for {\em stochastic}) by
\be \label{gas1}
     \omega^{}_s \; = \; \sum_{x\in\Gamma} \eta(x) \delta_x
\ee
where $\eta(x)$ is a random variable at site $x$ which takes only
the values 0 and 1, meaning empty or occupied.
We assume that these random variables are
independent of one another and identically distributed, i.e.\ they
constitute a (countable) family of i.i.d.\ random variables 
\cite{Bauer,Sinai}.
In fact, we parameterize the probability $P$ that $\eta(x)$ takes
the value 1 by a number $0 \leq p \leq 1$, i.e.\
\be \label{param}
       P(\{\eta(x)=1\}) \; = \; p \quad , \quad 
            \quad P(\{\eta(x)=0\}) \; = \; 1-p \, .
\ee
With this definition, $\omega^{}_s$ of (\ref{gas1}) describes scatterers
on the sites of a lattice, each single site being occupied with the
same independent probability $p$. Clearly, the mean value of each
random variable is $m^{}_1 = p$, the second moment is 
$m^{}_2 = p$ and the variance is thus $v = m^{}_2 - (m^{}_1)^2 = p (1-p)$.

Refering now to the strong law of large numbers \cite{Bauer,Sinai}, we can
deduce (details will be given below in a more general context) that,
almost surely,
\be
    \lim_{r\rightarrow\infty} \frac{1}{{\rm vol}(B_r(0))}
          \sum_{x\in\Gamma_r} \eta(x) \eta(x-y)
    \; = \; d \cdot p ( p + (1-p) \delta_{y,0}) \; ,
\ee
i.e.\ the limit exists and equals the right hand side
with probability one. Here, $\delta_{a,b}$ denotes Kronecker's delta. 
With this expression, one can calculate the new autocorrelation
to be, with probability one,
\be
    \gamma^{}_{\omega^{}_s} \; = \; p^2 \cdot \gamma^{}_{\omega}
          + d\cdot p(1-p)\cdot \delta_0 \, .
\ee
Fourier's transformation then gives
\be \label{gasdiff}
    \hat{\gamma}^{}_{\omega^{}_s} \; = \; p^2 \cdot \hat{\gamma}^{}_{\omega}
         + d \cdot p(1-p) \; ,
\ee
i.e.\ we retrieve the point part of the full lattice case, reduced
by a factor of $p^2$ as it should according to the reduced
density of points, plus a constant diffuse background which
is the absolutely continuous part of the diffraction. This term
is related to the entropy density $s = s(p)$,
\be
      s(p) \; = \; - p \log(p) - (1-p) \log(1-p) \, ,
\ee
which is a measure of the complexity of the ensemble of
point sets we are actually dealing with in this example. 
The presence of entropy is usually connected to continuous components
in the diffraction spectrum, and one can clearly see that $s(p)$
vanishes if, and only if, $p=0$ or $p=1$, i.e.\ iff the random
variables are sharp. Also, $s(p)$ is maximal at $p=1/2$, which corresponds
to the value of $p$ where the diffraction (\ref{gasdiff}) shows the
largest amount of white noise. Finally, in this case, we have an
essentially unique relationship between the entropy and the background
intensity, up to the symmetry $p \; \leftrightarrow \; (1-p)$.

Note that the continuous part vanishes if and only if the variance
of our random variable vanishes, i.e.\ if and only if $p=0$ or $p=1$.
Also, the pure point part vanishes if and only if the mean of the
random variable vanishes, i.e.\ if and only if $p=0$. We shall meet this
situation again later in a much more general context.

\subsection*{General setup}

The previous section should have served to get a feeling of what we are after,
and to introduce the type of notation we wish to apply. Let us now develop the
theory with more precision and in more generality. 
In what follows, we will only consider
uniformly discrete point sets $\Lambda\subset\RR^n$ here, 
i.e.\ points sets with the property that 
there is a positive radius $\varepsilon$ such that
each point $x\in\Lambda$ can be surrounded by an open ball of radius 
$\varepsilon$ that does not contain any point from $\Lambda$ other than $x$. 
With this assumption, the corresponding Dirac comb $\omega^{}_{\Lambda}$ defines 
a translationally bounded distribution (i.e., for each compact set
$K$, there is a constant $c_K$ so that for all $x \in \RR^n$,
$(\omega^{}_{\Lambda},\chi^{}_{K+x}) \;\leq \; c_K$, where $\chi^{}_S$ denotes the
characteristic function of a set $S$). This is sufficient, though certainly
not necessary, to make $\omega^{}_{\Lambda}$ a tempered distribution.
As before, we write
$\Lambda_r$ for the intersection $\Lambda\cap B_r(0)$. Now, we have
to tie this together with Fourier analysis.
\begin{defin} \label{weakdiff}
  Let $\Lambda$ be a uniformly discrete point set of (existing) natural density 
  $d  > 0$, and let $\omega = \omega^{}_{\Lambda}
  = \sum_{x\in\Lambda} \delta_x$ be its Dirac comb.
  We say that  $\Lambda$ has a natural {\em autocorrelation} if
\be \label{auto}
    \gamma^{}_{\omega} \; := \; \lim_{r\rightarrow\infty}
          \frac{1}{{\rm vol}(B_r(0))} \sum_{x,y\in\Lambda_r}
          \delta_{x-y}
\ee
exists as a limit in the weak topology (i.e., as a limit of tempered
distributions), and thus is a tempered distribution. Then, 
$\gamma^{}_{\omega}$ is called an {\em autocorrelation distribution} 
or simply an {\em autocorrelation} of $\Lambda$.
\end{defin}
Note that, if this situation applies, then $\gamma^{}_{\omega}$ is 
translation bounded ($\gamma^{}_{\omega}$ inherits this property from
$\omega$) and is a distribution of positive type, compare \cite{Hof}.
Let us briefly comment on the more general setup in terms of measures.
If we only had existence of an autocorrelation as a measure in the vague 
topology, translation boundedness would guarantee that it is actually
tempered -- so, in our context of uniformly discrete sets, the restriction 
to tempered distributions is reasonable.

Note that existence of an autocorrelation, as we have defined it here,
is specific to the type of region (in this case balls, as implied above by 
the attribute ``natural'') over which we compute our averages.
Replacing balls by other (convex) objects, centered at 0, the limit 
$r\rightarrow\infty$ (with $r$ the radius of inscribed balls, say) might give
a different answer or might not even exist. For the purposes of this article 
we do not need to deal with limiting processes over more than one type of
shape at the same time, so we will phrase
the arguments in terms of convergence based on sequences of balls. The arguments
for sequences based on other shapes work in the same way. However, for other
purposes it is important to specify {\em uniqueness} of the autocorrelation. 
For example, this can be done as follows:

Let $(C_n)_{n\in\NN}$ be a family of convex bodies, centered at 0, with the properties
that, as $n\rightarrow\infty$, the radius of the maximal inscribed balls tends to 
$\infty$ and the quotient of the radii of circum- and inscribed balls is bounded.
If 
\be \label{auto2}
    \gamma^{}_{\omega} \; := \; \lim_{n\rightarrow\infty}
          \frac{1}{{\rm vol}(C_n)} \sum_{x,y\in\Lambda\cap C_n}
          \delta_{x-y}
\ee
exists for each such sequence as a tempered distribution and is unique, we say
that $\Lambda$ has a {\em unique autocorrelation}, $\gamma^{}_{\omega}$.

In any case, a point set with a natural autocorrelation has a positive measure 
$\hat{\gamma}^{}_{\omega}$ as its Fourier transform (due to the Bochner-Schwartz
theorem \cite{RS}). It is this measure that desribes the diffraction 
\cite{Cowley,Hof}, and it is very natural that a positive measure shows up here:
after all, diffraction is all about the amount of intensity scattered into
a certain (measurable) reagion of space. Also, this positive measure
can now, according to Lebesgue's decomposition theorem, 
uniquely be decomposed into an absolutely continuous part (ac), a singular
continuous part (sc), and a pure point part (pp).  The pure point part
(usually called ``Bragg part'' in physics)
will always contain a trivial term of the form $d^2\cdot \delta_0$ where $d$ is
the (by assumption existing) density of $\Lambda$ per unit volume. This motivates 
\begin{defin}
  A point set $\Lambda$ with autocorrelation $\gamma^{}_{\omega}$
  is called {\em diffractive} (with respect to the convergence process
  adopted) if $(\hat{\gamma}^{}_{\omega})_{pp}$ is non-trivial,
  i.e.\ contains Dirac distributions different from $d^2\cdot
  \delta_0$.  $\Lambda$ is called {\em perfectly diffractive} or {\em
  pure point} if $\hat{\gamma}^{}_{\omega}$ has no continuous part at
  all.  \end{defin}

The simplest example of a perfectly diffractive point set is a lattice,
where the statement follows from Poisson's summation formula. Another
class of examples is given by regular model sets with sufficiently nice
windows (see the last section for more on this) or by extensions of them to 
certain limit-periodic or limit-quasiperiodic point
sets usually described by means of inflation \cite{Solomyak,BMS}. All these
examples are not only uniformly discrete, but also relatively dense, so
they are Delone sets. What is more, they are actually Meyer sets,
i.e.\ not only are they Delone but they have the additional property that 
their difference set, $\Lambda - \Lambda$, is also Delone. Note, however, 
that the Delone property is not
necessary for perfectly diffractive sets, as can be seen from the
example of the set of visible points of a lattice \cite{BaMo} which has
holes of arbitrary size (and this even with positive density)
and is thus neither Delone nor a density 0 deviation of one.
Note that removing or adding points
of density zero from a perfectly diffractive set does not change its
autocorrelation, and the set thus stays perfectly diffractive.

On the other hand, Meyer sets need not be perfectly diffractive, as can be seen
from the union of $2\ZZ$ with various subsets of $2\ZZ + 1$. This is always a
Meyer set, but one can easily construct cases with continuous components
(and positve entropy density).
This indicates that the class of Meyer sets, or even Delone sets, and the class
of perfectly diffractive sets are rather different, though they have some sets
in common. In general, perfectly diffractive sets will not be Delone, and
hence not Meyer. One interesting class of point sets in this context is that
of uniformly discrete sets $S$ with the extra property that $S-S$ is Delone,
or at least that $S-S$ is closed and discrete. They are the ones we shall
consider here.

\subsection*{Point sets with independent stochastic occupation of sites}

In this Section, we will develop an appropriate generalization of the
lattice gas (with i.i.d.\ random variables) to much more general point sets.
From now on, let $\Lambda$ be a uniformly discrete point set which 
has a {\em natural autocorrelation}. Let us also assume that $\Lambda$ is of 
{\em finite local complexity}, i.e., that $\Delta := \Lambda-\Lambda$ is 
discrete and closed, compare \cite{Lagarias} for a detailed discussion
in the context of Delone sets. Finite local complexity of a set $\Lambda$
implies that, for every radius $r>0$, there are, up to translations, only 
finitely many different configurations of points in a ball of radius $r$.
In particular, $\Lambda$ is uniformly discrete. This is so because
$\Delta$ discrete forces $0\in\Delta$ to be isolated, so different points 
in $\Lambda$ must have a uniform minimal distance from one another.

If $\omega = \sum_{x\in\Lambda}\delta_x$ is the Dirac comb of $\Lambda$, as 
usual, it is now certainly translation bounded, and 
the autocorrelation is given by Eq.~(\ref{auto}). We then have
\begin{equation}
    \gamma^{}_{\omega} \; = \;  \sum_{z\in\Delta} \nu(z) \delta_z
\end{equation}
where $\nu(z)$ is the {\em autocorrelation coefficient} at $z$, defined by
\begin{equation} \label{coeff1}
    \nu(z) \; = \; \lim_{r\rightarrow\infty} {1\over {\rm V}_r} 
      \sum_{\stackrel{\scriptstyle x,y \in \Lambda_r}
                     {\scriptstyle x-y = z}} 1 \, .
\end{equation}
Here, $\Lambda_r := \Lambda\cap B_r(0)$ as before, and 
${\rm V}_r := {\rm vol}(B_r(0))$. Note that, in order to establish the
existence of an autocorrelation, it is sufficient to show the existence of
the limits in (\ref{coeff1}), i.e.\ the existence of the coefficients, because
$\Delta$ discrete then implies existence of the autocorrelation as a measure
in the vague topology, and translation boundedness ensures temperedness,
see \cite{Hof} for further details.

Let us now turn to a stochastic ``lattice gas'' version of $\Lambda$. 
It is defined by the characteristic distribution 
\begin{equation}
     \omega^{}_s \; = \; \sum_{x\in\Lambda} \eta(x) \delta_x \, , 
\end{equation}
where $\eta(x)$ is a family of i.i.d.\ random variables taking the
values 0 and 1, parameterized as in Eq.~(\ref{param}), each with mean $p$ and 
variance $v=p(1-p)$.

We first address the question of the existence of the corresponding 
stochastic autocorrelation.
In analogy to Eq.~(\ref{coeff1}), we now have the coefficients
\be \label{coeff2}
\nu^{}_s(z) \; := \; \lim_{r\rightarrow\infty} {1\over {\rm V}_r} 
      \sum_{\stackrel{\scriptstyle x,y \in \Lambda_r}
                     {\scriptstyle x-y = z}} \eta(x)\eta(y) 
          \;  = \;
\lim_{r\rightarrow\infty} {1\over {\rm V}_r} 
      \sum_{\scriptstyle x,x-z \in \Lambda_r} \eta(x)\eta(x-z)  .
\ee

We will show that under mild assumptions these coefficients exist,
at least in a probabilistic sense.
In order to do this, we need to be able to decompose
the sum involved in $\nu^{}_s(z)$ into two parts, because the various
terms in the sum of (\ref{coeff2}) are still random variables, but
not necessarily independent ones any more. Fix $z\in \RR^n$. Define
\be
     S(z) \; := \; \{x \mid x , x-z \in \Lambda \}
\ee
and its restricted version $S(z,r)$, where the $x, x-z$ appearing
in the definition are required to lie in $\Lambda_r$. We
distribute the points of $S(z)$ (and, by proper restriction, also those of
$S(z,r)$) into {\em two} sets $S(z)^{(0)}$ and $S(z)^{(1)}$. This may be
done in an arbitrary fashion, subject only to the two conditions that \newline
(1) if $x, x-z$ both lie in $S(z)$, then they are {\em not} 
    in the same $S(z)^{(i)}$,  and  \newline
(2) the two sets $S(z)^{(i)}$ have well-defined densities:
\be \label{sepa}
       \nu^{(i)}(z) \; := \; \lim_{r \to \infty} \frac{1}{V_r}
         \sum_{x \in S(z,r)^{(i)}} 1 \, .
\ee
Evidently, (\ref{sepa}) implies $\nu(z) = \nu^{(0)}(z) + \nu^{(1)}(z)$.
Let us say that the set $\Lambda$ can be {\em decoupled} if, for every
$z\in \Delta$, we can find such a partition. 

We note that there may be many ways of decoupling a set $\Lambda$.
For lattices, for example, we can take each line of points 
$x+\ZZ z$ and distribute
it into the two subsets according to whether the coefficient of $z$ is
even or odd. For aperiodic model sets (see definitions
below), each of the sets 
$\{x \mid x+z, x - kz \notin \Lambda ~; ~ x, x-z, \dots, x -(k-1)z \in 
\Lambda\}$ is finite, of bounded length $k$ and, for each $k$,
the set of such strings has a definite density. We can place
the points $x, x-z, x-2z, \dots $ alternately into $S(z)^{(0)}$
and $S(z)^{(1)}$ and again obtain sets with well-defined density this way;
see \cite{Hof95} for a very similar approach to thermal fluctuations
which establishes the usual form of the Debye-Waller factor for
essentially the same kind of structures that we are dealing with here.

The decoupling property is some kind of ergodicity assumption. 
It is certainly fulfilled for point
sets with uniform frequencies of all finite patches (as is the case for
usual model sets), but it is more general than this. In particular, it
is still valid for objects such as the pinwheel tiling, compare the brief
discussion in \cite{Hof95}. At present, we do not know any equivalent
characterization simpler than that given above, which is very much designed
for its (technical) purpose.

\begin{prop}
Let $\Lambda$ be a point set of finite local complexity which has a natural 
autocorrelation. Assume further that the set $\Lambda$ can be decoupled.
Then each coefficient of the stochastic autocorrelation 
(i.e.\ the corresponding limit) exists with probability 1, and is given by
\begin{equation}
        \nu^{}_s(z) \; = \; \nu(z) \cdot 
        \left( (m_1)^2 + (m_2 - (m_1)^2) \delta_{z,0} \right) \, 
\end{equation}
where $m_1 (= p)$ is the common mean of the i.i.d.\ random variables
$\eta(x)$ and $m_2$ is their common second moment. 
\end{prop}
{\sc Proof}: This is an application of the strong law of large numbers.
Forming a sequence of random variables out of a family 
$(\eta(x))_{x\in\Lambda}$ etc.\ is rather canonical. 
Since $\Lambda$ is uniformly discrete, we number
the $\eta(x)$ with $x$ in finite sets $\Lambda_r$ for increasing $r$. Each
such sequence, by the general assumptions made, is a sequence that conforms to
the strong law of large numbers.

Let us consider $z=0$ first. Here, the relevant random variable is actually 
$\eta(x)^2$, with mean $m_2$, the second moment of $\eta(x)$. These variables 
are independent and, almost surely,
$$\lim_{r\rightarrow\infty} {1\over {\rm V}_r}
     \sum_{\stackrel{\scriptstyle x\in\Lambda_r}{\scriptstyle x-z\in\Lambda}}
     \eta(x)^2 \; = \; d \cdot m_2 $$
where $d=\nu(0)$ is the (existing) natural density of $\Lambda$.

Next, let $z\neq 0$, $z\in\Delta$, be arbitrary, but fixed. If $\nu(z)=0$,
also $\nu^{}_s(z)=0$, and our assertion is trivial. So, assume $\nu(z)>0$, which
means that the density of points $x\in\Lambda$, such that also $x-z\in\Lambda$,
exists and is positive. For each such $x$, $\eta(x)\eta(x-z)$ is a random variable
with mean $(m_1)^2$, where $m_1 = p$ is the (identical) mean of all random
variables $\eta(y)$ involved. We now have to consider
$$\lim_{r\rightarrow\infty} {1\over {\rm V}_r}
     \sum_{\stackrel{\scriptstyle x\in\Lambda_r}{\scriptstyle x-z\in\Lambda_r}}
     \eta(x)\eta(x-z) \,.$$
This sum has only non-negative terms and decomposes as two sums:
$$
\nu^{}_s(z) \;= \; \lim_{r\rightarrow\infty} {1\over {\rm V}_r}
     \sum_{\scriptstyle x\in S(z,r)^{(0)}}
     \eta(x)\eta(x-z) \; + \; 
\lim_{r\rightarrow\infty} {1\over {\rm V}_r}
     \sum_{\scriptstyle x\in S(z,r)^{(1)}}
     \eta(x)\eta(x-z) \, .
$$
Now each of the two sums is an averaged sum over a set of {\em independent}
random variables. Hence, by the strong law, we get almost sure
convergence to
\be
 \nu^{}_s(z) \; = \;  \nu^{(0)}(z) \, m_1^2 + \nu^{(1)}(z) \, m_1^2 
      \; = \; \nu(z) \, m_1^2 
\ee
because the mean of each random variable $\eta(x)\eta(x-z)$ is $m_1^2$.
Together with the first step, this establishes our claim. $\square$

The autocorrelation $\gamma^{}_{\omega^{}_s}$ of $\omega^{}_s$ is defined
as the distribution whose value on any test function $\phi$ is
\begin{equation}
(\gamma^{}_{\omega^{}_s}, \phi) \;= \; \lim_{r\rightarrow\infty} {1\over {\rm V}_r}
      \sum_{x,y \in \Lambda_r} \eta(x)\eta(y)\phi(x-y)\, . 
\end{equation}
Although we know this already from the above abstract arguments, it might 
be instructive to check explicitly that $\gamma^{}_{\omega^{}_s}$ is indeed a 
tempered distribution, at least in the sense of almost sure convergence.
First, let $\phi$ be a $C^\infty$ function of compact support,
lying in the ball $B_s(0)$ of radius $s$. Then
\begin{equation}
  ( \gamma^{}_{\omega^{}_s}, \phi ) \; = \;
\lim_{r\rightarrow\infty} 
    \sum_{z\in\Delta}{1\over {\rm V}_r} \left( 
  \sum_{\stackrel{\scriptstyle x,y \in \Lambda_r}{\scriptstyle x-y =
    z}} \eta(x)\eta(y)\right) \phi(z) \, .
\end{equation}
This limit exists because in reality the outer sum is over the
finite set $\Delta_s$
and for $r>\!>s$ we have  $\Delta_s \subset \Lambda_r - \Lambda_r$.
Thus, as $r \rightarrow \infty$, the sum converges, almost surely, to
\begin{equation}
\sum_{z\in\Delta} \nu^{}_s(z) \phi(z) \; .
\end{equation}

Now if $\phi \in {\cal S}$, the space of Schwartz functions,
and $\{\phi_i\}$ is a sequence of $C^\infty$-functions of compact 
support that converge to $\phi$ in the standard topology of ${\cal S}$, then 
\begin{equation}
 \sum_{z\in\Delta} \nu^{}_s(z)\phi_i(z)  
           \; = \; (m_1)^2 \sum_{z\in\Delta} \nu(z)\phi_i(z) 
                   + (m_2 - (m_1)^2) \nu(0) \phi_i(0)\, ,
\end{equation}
and the latter converges in $i$ to
\begin{equation}
(m_1)^2 \sum_{z\in\Delta} \nu(z)\phi(z) 
 + (m_2 - (m_1)^2) \nu(0) \phi(0)  ,
\end{equation}
by our assumptions on the existence
of the autocorrelation density of $\Lambda$.
 
Let us summarize these findings as follows.
\begin{theorem}
Let $\Lambda$ be a point set of finite local complexity which has a natural 
autocorrelation and density $d$. Suppose that $\Lambda$ can be decoupled. 
Then, the autocorrelation of $\Lambda$ and that of its stochastic version 
are, with probability one, related by
\begin{equation}
        \gamma^{}_{\omega^{}_s} \; = \; 
          (m_1)^2 \, \gamma^{}_{\omega} + d\, (m_2 - (m_1)^2)\, \delta_0 \, .
\end{equation}
As a consequence, their Fourier transforms fulfil
\begin{equation} \label{stochdiff}
       \hat{\gamma}^{}_{\omega^{}_s} \; = \; 
          (m_1)^2 \, \hat{\gamma}^{}_{\omega} + d\, (m_2 - (m_1)^2) \, .
\end{equation}
\end{theorem}

So, the stochastic version has the same `main' part of the diffraction,
multiplied by a factor of $(m_1)^2$ (hence vanishing if and only if
the mean of the joint probability distribution is $0$) plus an extra absolutely
continuous part that is constant and represents the `white noise' of the
uncorrelated random processes. The constant is essentially given by the variance 
of the joint distribution, and  thus this part vanishes if and only if the i.i.d.\ 
random variables are all sharp. The interpretation, and also the connection
with the entropy density, is thus the same as in the lattice case, as expected.

At this point, generalizations are rather obvious, and we just want to mention
a few. First of all, it is by no means essential to restrict to the particular
types of random variables that we have just discussed. 
Here we were motivated by the idea of a lattice 
gas and its generalization to uniformly discrete point sets, 
but we can also  think of any other (non-negative) i.i.d.\ random variable
with (existing) mean $m_1$ and second moment $m_2$ (so, the variance would be
$v = m_2 - (m_1)^2$). This does not change the result, and would correspond
to a situation where we place, at each point $x$ of the set $\Lambda$, a scatterer
of random strength $\eta(x)$. Again, we get the result of Theorem 1.

Note also that at no point did we need to assume that $\hat{\gamma}^{}_{\omega}$
was pure point. This is not necessary, indeed, and the result of Theorem 1
also applies to situations where $\hat{\gamma}^{}_{\omega}$ is singular
continuous, absolutely continuous, or of mixed type. This same situation is
met in Hof's treatment of thermal fluctuations \cite{Hof95}.

\subsection*{Applications to lattices and model sets}

The obvious first application is to {\em lattices}. This results in a rigorous 
derivation of what we described in Section 2. The diffraction from a lattice 
gas, with i.i.d.\ random variables
for the strength of the Dirac distributions at the lattice points, shows a
point part that is the one from the lattice itself, reduced in intensity, plus
a homogeneous diffuse background. 

Another application is to characteristic decorations 
on tilings\footnote{We call a point set of finite local complexity a
characteristic decoration of a locally finite tiling if they are locally
equivalent, i.e.\ if both objects represent the same MLD-class, see \cite{BS}
for details.}   that are obtained
by a primitive substitution rule. Here, it was shown \cite{LP} that the
autocorrelation is unique, and convergence of its coefficients is even uniform.
In general, the Fourier transform will not be pure point, see \cite{Solomyak}
for a more detailed discussion. We also refer to \cite{Hof95} for a brief
discussion of the decoupling property in situations without finite
local complexity such as the pinwheel tilings of the plane. 

Lattice gas versions of {\em model sets} provide another class of examples,
of rather recent interest. Recall that a model set 
\cite{Moody,MartinFields} is defined via
projection onto $\RR^n$ of a lattice in some higher dimensional space,
or, more generally, in some locally compact Abelian group. More
precisely, it is assumed that $G = \RR^n \times H$ is a locally compact
Abelian group and that $D$ is a lattice in $G$. Thus $D$ is a discrete
subgroup of $G$ for which the quotient space $G/D$ is compact. Further we
assume that the projection $\pi^{}_1(D)$ of $D$ into $\RR^n$ is 
{\em injective} and
its projection $\pi^{}_2(D)$ into $H$ is {\em dense}, 
where $\pi^{}_1$ and $\pi^{}_2$ denote
the canonical projections. The resulting set is aperiodic, i.e. has no
translational symmetries, if and only if $\pi^{}_2$ is injective on $D$.
The most common examples take $H = \RR^m$ for some $m$. 
In any case, define the composite map 
$^* := \pi^{}_2 \circ \pi^{}_1|_D^{-1} : \pi^{}_1(D) \longrightarrow H$. 
Then for any set $\Omega \subset H$ with
nonempty interior and compact closure,  we have the {\em model set}
\begin{equation}
       \Lambda \; = \; \{ x \in \pi^{}_1(D) \mid x^* \in \Omega \} \, .
\end{equation} 

Provided that the boundary of $\Omega$ has measure
$0$ (with respect to the Haar measure $\mu$ of $H$), the density of such a 
set exists uniformly and is given by $\mu(\Omega)/{\rm vol}(D)$. Here, 
${\rm vol}(D)$ is the volume of any fundamental domain
for $D$ in $G$, the volume taken relative to the product measure on $G$
derived from the Lebesgue measure on $\RR^n$ and the Haar measure $\mu$ on $H$
\cite{Martin, MartinFields}. Such a model set is a Meyer set, i.e.\ both
$\Lambda$ and $\Delta = \Lambda - \Lambda$ are Delone. Also, $\Lambda$
is perfectly diffractive \cite{Martin98}, and the obvious lattice gas 
version of it, with i.i.d.\ random variables attached to each position, 
falls under our Theorem 1. 

A large number of well-known point sets can
be interpreted in this setting, including the Fibonacci 
and many other chains, the
vertex sets of various planar tilings (such as the Ammann-Beenker,
the Penrose, the T\"ubingen triangle tiling etc.) or of tilings in
higher dimensions (such as the various icosahedral examples in 3D or
the Elser-Sloane quasicrystal in 4D). But even decorations of the
chair tiling and other limit-periodic and limit-quasiperiodic
structures fall under this class, see \cite{BMS} for details. 
So, for all these cases, we have
\begin{coro}
If $\Lambda$ is a model set as described above, it fulfils the conditions
of Theorem 1, and the diffraction of the stochastic versus the
deterministic Dirac comb is, almost surely, given by 
Eq.~(\ref{stochdiff}).
\end{coro}

\subsection*{Beyond identical distribution}

So far, we have restricted our attention to the case of i.i.d.\ variables.
We will now broaden our point of view to the situation where the random
variables are still independent, but not necessarily identically distributed
any more. Before we give proper definitions, let us have another look at
the lattice gas. Above, we compared the deterministic Dirac comb
$\omega = \sum_{x\in\Gamma}\delta_x$ with the stochastic one,
$\omega^{}_s = \sum_{x\in\Gamma}\eta(x)\delta_x$, where $\eta(x)$ were i.i.d.\
random variables of common mean $m^{}_1$. This led to Theorem 1.

Alternatively, consider now the deterministic, but weighted distribution
\begin{equation}
      \omega^{}_a \; := \; \sum_{x\in\Gamma} m^{}_1 \delta_x \, .
\end{equation}
Clearly, $\omega^{}_a$ has an autocorrelation if $\omega$ itself does, and we get
\begin{equation}
       \gamma^{}_{\omega^{}_a} \; = \; (m^{}_1)^2 \cdot \gamma^{}_{\omega}
\end{equation}
and the result of Theorem 1 may be restated as
\begin{equation}
      \hat{\gamma}^{}_{\omega^{}_s} \; = \; \hat{\gamma}^{}_{\omega^{}_a}
                                         + d\cdot(m^{}_2 - (m^{}_1)^2)
\end{equation}
which holds almost surely. 

This indicates how we have to generalize our previous findings properly.
Let $\Lambda$ again be a uniformly discrete set of finite local complexity
(i.e.\ $\Delta = \Lambda - \Lambda$ discrete and closed), and suppose that 
$\Lambda$ can be decoupled in the sense described above. Furthermore,
let $(\eta(x))_{x\in\Lambda}$ be a family of independent random variables
with non-negative means $m^{}_1(x)$ which are bounded from above and 
with bounded variances $v(x)$, $v(x)\leq c$, say. 
Under these assumptions, this family conforms to the strong
law of large numbers. This can be seen as follows.
Let $(\eta_m)_{m\in\NN}$ be any sequence made
from the random variables $\eta(x)$, e.g.\ by numbering the points of $\Lambda$
in balls of growing radius. We then obtain
\begin{equation}
    \sum_{m=1}^{\infty} \frac{v(\eta_m)}{m^2}
    \; \leq \; c \, \sum_{m=1}^{\infty} {1\over m^2}
    \; = \; c \cdot \zeta(2) \; = \; \frac{c \pi^2}{6} \; < \infty \, .
\end{equation}
The assertion now follows from Kolmogorov's criterion, see \cite{Bauer}
or \cite[Theorem 12.3]{Sinai}, and we can 
continue to develop the appropriate analogue of Theorem 1.

To this end, let us now compare the two distributions
\begin{equation}
   \omega^{}_a \; = \; \sum_{x\in\Lambda} m^{}_1(x) \delta_x \, ,
\end{equation}
which may be thought of as a toy model for an arrangement of different atoms
(hence the suffix $a$), and
\begin{equation}
     \omega^{}_s \; = \; \sum_{x\in\Lambda} \eta(x) \delta_x \, ,
\end{equation} 
the former being deterministic and the latter probabilistic. We can
now formulate the appropriate theorem for this situation.
\begin{theorem}
  Let $\Lambda$ be a set of finite local complexity that can be decoupled, and
  let $(\eta(x))_{x\in\Lambda}$ be a family of independent random variables
  with non-negative means $m^{}_1(x)$ (bounded from above) and bounded
  variances $v(x)$ whose average is assumed to exist,
\begin{equation}
    \overline{v} \; = \; \lim_{r\rightarrow\infty} \frac{1}{|\Lambda_r|}
                    \sum_{x\in\Lambda_r} v(x) \, .
\end{equation} 
  If $\omega^{}_a$ has a natural autocorrelation, in the sense
  we used this term above, then $\omega^{}_s$ also possesses, almost surely, a 
  natural autocorrelation, namely, with $d=\mbox{dens}(\Lambda)$,
\begin{equation}
    \gamma^{}_{\omega^{}_s} \; = \; 
              \gamma^{}_{\omega^{}_a} + d\, \overline{v} \, \delta_0
\end{equation}
  and its Fourier transform reads
\begin{equation}
   \hat{\gamma}^{}_{\omega^{}_s} \; = \; 
              \hat{\gamma}^{}_{\omega^{}_a} + d\, \overline{v} \, .
\end{equation}
\end{theorem}
The proof is very similar to the one given above and need not be repeated.

This theorem is a little less explicit than the previous one, and one can
see the potential extra complication from the following simple example.
Consider $\Lambda = \ZZ$ and independent random variables $\eta(m)$ with
values in $\{0,1\}$ and parametrization
\begin{equation}
    P(\{\eta(m) = 1\}) \; = \; \cases{p, & $m$ even,\cr
                                   q, & $m$ odd, }
 \end{equation}
where $0\leq p,q\leq 1$. Here, $\omega^{}_a = q \sum_{x\in\ZZ}\delta_x
+ (p-q) \sum_{x\in 2\ZZ}\delta_x$ which clearly has a unique
autocorrelation $\gamma^{}_{\omega^{}_a}$, with Fourier transform
\begin{equation}
    \hat{\gamma}^{}_{\omega^{}_a} \; = \; 
         \frac{(p+q)^2}{4} \sum_{y\in\ZZ} \delta_y +
         \frac{(p-q)^2}{4} \sum_{y\in\ZZ+{1\over 2}} \delta_y \, .
\end{equation}
So, the diffraction spectrum depends on the values of $p$ and $q$, and the
second term on the right hand side vanishes for $p=q$.

The corresponding stochastic version, $\omega^{}_s$, reflects this and produces 
the same point diffraction, plus a constant diffuse background
(``white noise''), i.e.\ by application of Theorem 2 we have, almost surely,
\begin{equation}
     \hat{\gamma}^{}_{\omega^{}_s} \; = \; \hat{\gamma}^{}_{\omega^{}_a} 
             + {1\over 2} \, [p(1-p) + q(1-q)] \, .
\end{equation}
The entropy density of this little example is immediate:
\begin{equation}
      s \; = \; - {1\over 2} \, [p\log(p) + (1-p)\log(1-p)
                       + q\log(q) + (1-q)\log(1-q) ] \, .
\end{equation}

Note that, in general, a perfectly diffractive point set $\Lambda$ together with a 
family of independent random variables $\eta(x)$ is not enough to apply 
Theorem 2; we 
really have to know that not only $\gamma^{}_{\omega}$ but also 
$\gamma^{}_{\omega^{}_a}$ exists. This is a rather subtle (and non-constructive) 
set of conditions upon the means of the random variables. There is one situation
where we can escape this extra complication: if the random variables are
distributed statistically, i.e.\ in such a way that their means $m^{}_1(x)$
are themselves the result of a stationary Bernoulli process, we are back to the 
situation of Theorem 1, which may then be applied with
\begin{equation}
        m^{}_1 \; = \; \lim_{r\rightarrow\infty} \frac{1}{|\Lambda_r|}
           \sum_{x\in\Lambda_r} m^{}_1(x) \, ,        
\end{equation}
provided this limit exists.

\subsection*{Further examples: weighted model sets}

Let us now come back to the situation of a model set $\Lambda=\Lambda(\Omega)$, 
as described above. Assume that we have a family of independent
random variables parametrized by the points $x^*$ of the window $\Omega$.
Suppose that $p(x^*)$ is a continuous function on $\Omega$, with
values in $[0,1]$. Then, $\omega^{}_a = \sum_{x\in\Lambda}p(x^*)\delta_x$
is perfectly diffractive (this follows from a slight modification of
the arguments given in \cite{Hof} by means of an application of
Weierstrass' approximation theorem). Explicitly, one has
\begin{equation}
   \hat{\gamma}^{}_{\omega^{}_a} \; = \;
          \sum_{k\in\pi^{}_1(D^{\circ})} |a(k)|^2 \, \delta_k
\end{equation}
where $D^{\circ}$ is the dual lattice of $D$ and the amplitudes are given 
by \cite{Hof}
\begin{equation}
       a(k) \; = \; \frac{d}{{\rm vol}(\Omega)} \,
            \int_{\Omega} e^{2\pi i k^* \cdot x^*} p(x^*) d\mu(x^*) 
            \; = \; \frac{d\cdot\hat{p}(-k^*)}{{\rm vol}(\Omega)}  
\end{equation}
where $d$ denotes, as before, the density of the model set $\Lambda$.

Let us turn to the stochastic counterpart
\begin{equation}
     \omega^{}_s \; = \; \sum_{x\in\Lambda} \eta(x) \delta_x \, ,
\end{equation}
where $\eta(x)$ is the random variable that decides whether $x$ is occupied
or not. Let us define it as follows
\begin{equation}
     P(\{\eta(x) = 1\}) \; = \; p(x^*) \; , \;\;
     P(\{\eta(x) = 0\}) \; = \; 1 - p(x^*) \; .
\end{equation}
Observe that $\eta(x)$ has mean $p(x^*)$ and variance $p(x^*)(1-p(x^*))$, the
latter being bounded by $1/4$. So, by Kolmogorov's criterion, this family of
random variables conforms to the strong law of large numbers.

Let us see whether the mean of the variances exists.
We note first that we have the mean occupancy per point of $\Lambda$ as
\begin{equation}
        \overline{p} \; = \; \lim_{r\to \infty} 
       {1\over |\Lambda_r|} \sum_{x \in \Lambda_r} p(x^*) \, .
\end{equation}
Due to the fact that the projection into internal space $H$ is 
uniform and the fact that $p$ is continuous, it is possible to
use Weyl's theory of uniformly distributed sets \cite{Kuip} to show 
that this limit indeed exists and is given by 
\begin{equation}
   \overline{p} \; = \; \frac{1}{{\rm vol}(\Omega)}\int_{\Omega} p(y) d\mu(y)\, .
\end{equation}
In the same way, we can also calculate the averaged variance as
\begin{equation}
 \overline{v} \; = \; \frac{1}{{\rm vol}(\Omega)} \int_{\Omega} p(y)^2 d\mu(y) 
                     - \overline{p}^2 \, .
\end{equation}
So, we can apply Theorem 2 and obtain, with probability one,
\begin{equation}
     \hat{\gamma}^{}_{\omega^{}_s} \; = \; \hat{\gamma}^{}_{\omega^{}_a} 
             + \, d \,\overline{v} \, .
\end{equation}

The resulting set also has a positive entropy density. Clearly, for a single
point, this is
\begin{equation}
    s(x) \; = \; - p(x^*) \log(p(x^*)) - (1-p(x^*))\log(1-p(x^*))
\end{equation}
and we would be interested in the quantity
\begin{equation}
    \overline{s} \; := \; \lim_{r\rightarrow\infty} \frac{1}{|\Lambda_r|}
               \sum_{x\in\Lambda_r} s(x) \, ,
\end{equation}
provided this limit exists. Again, this follows from the
uniform distribution of the points of a model set 
and, using Weyl's lemma, we obtain
\begin{equation}
    \overline{s} \; = \; - \frac{1}{{\rm vol}(\Omega)} \int_{\Omega} 
              [p(y) \log(p(y)) + (1-p(y))\log(1-p(y))] \, d\mu(y) \, .
\end{equation}

As a relevant example, let us consider a special function $p(x^*)$,
namely one that reflects the inflation structure of a given model set and is
related to the recently investigated invariant densities on them 
\cite{BM97.1,BM97}.
Assume that $G = \RR^n \times \RR^m$ and suppose that 
$\Lambda = \{x \in \pi_1(D) \mid x^* \in \Omega \}$. We are interested here
in the situation in which $\Lambda$ admits self-similarities of the form 
\begin{equation}
      t_{Q,v} \; : \quad x \; \mapsto \; Qx +v \; ,
\end{equation}
where $Q$ is an inflational linear map, i.e.\ a rotation followed by a scalar 
inflation. We call such self-similarities $Q$-{\em inflations}. 
Remarkably, for fixed $Q$, the set 
$$T := \{v \in \RR^n \mid t_{Q,v}\Lambda\subset\Lambda\}$$ 
is itself a model set. In this situation, there
is a unique absolutely continuous probability measure $p = p^{}_Q$ 
supported on 
$\Omega$ which is invariant under the set of $Q$-inflations in the sense that
\begin{equation}
     p \; = \; \lim_{s \to \infty} \frac{1}{ |T_s| }
               \sum_{v\in T_s} t^*_v\cdot p ~,
\end{equation}
where $t^*_v$ is the induced mapping in internal space and 
$(t^*_v\cdot p) (y) := p((t^*_v)^{-1}y)$.

The corresponding stochastic model set with site occupancy probability 
$P(\{\eta(x)=1\})$,  is likewise invariant in the sense that 
\begin{equation}
    P(\{\eta(x) =1\}) \; = \; \lim_{s\to\infty}\frac{|{\rm det}(Q)|}{V_s} \,
        \sum_{v \in T_s} \; \sum_{y \in t_{Q,v}^{-1} x \cap \Lambda} 
              P(\{\eta(y) =1\}) \, .
\end{equation} 
Such invariant densities in internal space are supported on the
window and typically display bell-shaped form.
We refer the reader to \cite{BM97.1,BM97,BM98} for more details on this
and for various examples.

Since this is a special case of the general situation met above, the measure
\begin{equation}
    \omega^{}_s \; := \;  \sum_{x \in \Lambda} \eta(x)\delta_x
\end{equation}
is (almost surely) diffractive with the {\em same} pure point
part  $\omega^{}_a$, since
\begin{equation}
\lim_{r\to\infty} \frac{1}{V_r}\sum_{x, x-z \in \Lambda_r}
\eta(x)\eta(x-z)  \; = \;
\lim_{r\to\infty} \frac{1}{V_r}\sum_{x, x-z \in \Lambda_r}
p(x^*) p(x^* - z^*) \, . 
\end{equation} 

It is an interesting
feature of this situation that the probability distributions $p_{Q^k}$,
as $k\rightarrow\infty$, tend towards the constant distribution on $\Omega$.

\subsection*{Concluding remarks}

The analysis of diffraction from point sets with stochastic occupation of
sites, or with random scattering strength on the sites, can be developed in a rather
general setting which goes considerably beyond the lattice situation.
It was the aim of this contribution to outline some of the methods needed.
For related aspects, we also recommend Hof's treatment of thermal fluctuations
\cite{Hof95}.

One concrete reason to look into this type of problem stems from the discussion
of quasicrystalline order and the evidence of stochastic elements in it
\cite{JB,JRB}. Based upon the random tiling scenario, one would expect the
$*$-image (lift) of a ``real world'' point set (e.g.\ one obtained from a
tiling overlay of a high resolution electron micrograph) to show a Gaussian
shaped distribution or at least a bell shaped curve with maybe a somewhat
flatter centre -- in contrast to the uniform distribution obtained from
a perfect model set. 

Since such bell shaped distributions have been observed and appear to be
rather typical \cite{JRB}, it is an important question to what extent they really
support the random tiling picture. In other words: are there alternatives
to explain such profiles? One is provided by the stochastic occupation of a
model set, if we start from an invariant density on the window that resembles
such a bell curve, see \cite{BM97.1}, and \cite{BM98} for the specific
example of the Penrose tiling and invariant densities attached to it. 
A $*$-image of a finite patch its stochastic point set realization would 
reproduce the bell shaped invariant density profiles.

We do not claim that this is enough to establish this simple stochastic
approach as a real alternative -- there are various other objectives to be met,
such as width of the profile as a function of the patch size,
structured diffuse scattering background (other than white noise),
or maximization of entropy
as a function of suitable parameters for the non-crystallographic phase.
We believe, nevertheless, that there are interesting possibilities along
the lines presented here, and it would be nice to find (solvable) examples
with extra correlations that appear more realistic in the sense mentioned.

One first step in this direction is the calculation of the diffraction of
a stochastic version of $\ZZ$ with random variables that stem from a 
stationary ergodic Markov system, as is well known in the literature
\cite{Welberry}. This results in a pure point part which is
that of $\ZZ$, reduced in intensity, plus an absolutely continuous background
that now shows a structure, i.e.\ that is no longer white noise. We hope to
report on proper generalizations of this scenario soon, and also on an
extension to the diffraction theory of random tilings.

\subsection*{Acknowledgements}
It is our pleasure to thank Martin Schlottmann for several clarifying discussions
and Vladimir Rittenberg for helpful advise on the manuscript. This work was
supported by the German Science Foundation (DFG) and the Natural Science and
Engineering Research Council of Canada (NSERC).

\end{document}